\documentclass[aps,pra,showpacs,amsmath,amssymb,amsfonts,twocolumn,superscriptaddress,floatfix,footinbib]{revtex4}
  \usepackage{psfig}
  \usepackage{graphicx}

\newcommand{\p} {\prime}

\newcommand{\bgar}{\begin{eqnarray}}
\newcommand{\enar}{\end{eqnarray}}
\newcommand{\eq}[1]{(\ref{eq:#1})}
\newcommand{\eps}{\epsilon}
\newcommand{\ket}[1]{ \left. | #1 \right\rangle }

\newcommand{\eqname}[1]{\label{eq:#1}}

\newcommand{\El}{{{\bf E}}}

\newcommand{\J}{ {\bf J}}

\newcommand{\kk}{ {\bf k}}

\newcommand{\vel}{ {\bf v}}
\newcommand{\x}{ {\bf x}}
\newcommand{\Pv}{ {\bf P}}

\begin{document}
 \title{\bf In--situ velocity imaging of ultracold atoms using
slow--light.}

 \author{M. Artoni}
 \affiliation{INFM, European Laboratory for non-Linear Spectroscopy,
 Largo E.\ Fermi 2, 50125 Florence, Italy. }
 
 \author{I. Carusotto}
 \affiliation{Laboratoire Kastler Brossel, \'Ecole Normale
Sup\'erieure, 24 rue Lhomond, 75231 Paris Cedex 05, France}
 
\begin{abstract}
The optical response of a moving medium suitably driven into a slow--light propagation regime strongly depends on its
velocity. This effect can be used to devise a novel scheme for imaging ultraslow velocity fields. The scheme turns
out to be particularly amenable to study \textit{in--situ} the dynamics of collective and topological excitations
of a trapped Bose-Einstein condensate. We illustrate the  advantages of using slow--light imaging specifically for
sloshing oscillations and bent vortices in a stirred condensate.
\end{abstract}

 \date{\today}
 
 \maketitle

The dynamical properties of ultracold alkali atoms and Bose-Einstein condensates are
almost without exception inferred by
optical methods.
In most cases, these enable one to retrieve the atomic spatial density profile by using either
absorption or dispersive imaging
techniques. Because absorption is
followed by spontaneous emission and hence heating of the sample,
\textit{absorption imaging} is intrinsically
invasive. \textit{Dispersive imaging}, on the other hand, does not involve
much heating.
The density profile can in this case be reconstructed non invasively by measuring the phase--shift profile of the
transmitted imaging beam~\cite{dark-ground-ketterle,disp-ketterle,polar-hulet}. Yet, diffraction effects make
these techniques rather inappropriate when {\em in--situ} imaging has to be performed on structures that are
smaller than the wavelength of the imaging beam. For instance, a preliminary
ballistic expansion to enlarge the sample size is generally
required before images of a vortex in stirred Bose--Einstein
condensates can actually be taken~\cite{slicing,DalibBentVort,CornellVort}.

In--situ imaging is however necessary when real--time observations of a sample dynamics need to be made and we
here devise a new in-situ imaging scheme. This relies on the ultraslow propagation of the light imaging
beam~\cite{EIT} and enables one to image ultraslow velocity fields. In a regime in which the light group velocity
can be made to drop down to the m/s range~\cite{SlowVg,hemmer,ulf-general}, the optical response of a sample is
found to depend so strongly on its velocity that an imaging light beam may become sensitive enough to probe very
slow velocity fields of the sample.
At the same time, the strong quenching of absorption~\cite{EIT} that is typically observed in a regime of
slow--light propagation enables one to minimize the number of absorbed photons making the scheme inherently
non--invasive.

We will illustrate the physics underlying slow-light imaging while discussing two examples of
dynamical excitations of actual experimental interest in Bose-Einstein condensates. In one case, the
center--of--mass velocity of the entire cloud is imaged by measuring the lateral shift of a narrow probe beam
transversally propagating across the moving sample of atoms~\cite{FresTr}. In the second case, the internal
velocity field of the atomic cloud is imaged by measuring the phase-shift accumulated by the probe while crossing
the cloud. The spatial profile of the phase shift is shown to be proportional to the column integral of the {\em
atom current density} making this second scheme well suited to the {\em in--situ} imaging of topological
excitations such as vortices~\cite{stringari,Vortices}. In particular,
the phase shift profile
obtained by using slow-light turns out to be much less affected by
those diffraction effects which prevented a dispersive imaging of the
vortex core. 
Moreover, as the probe is now sensitive to the current density rather than to the
density itself, the vortex is not hidden by the surrounding stationary mass of fluid.

The dielectric function of a stationary sample of three-level atoms
driven into a lambda EIT--configuration by
a coupling beam of frequency $\omega_{c}$ and Rabi frequency $\Omega_{c}$
has the form~\footnote{This expression refers to the usual
single-atom model of the atomic
polarizability~\cite{EIT}. Work to include the finite optical
density of the medium as well as effects of
quantized atomic motion is under way.}:
\begin{equation}
\epsilon(\x,\omega_p)=1+\frac{4\pi f
N(\x)}{\omega_e-\omega_p-\frac{i\gamma_e}{2}-\frac{\Omega_c^2}{\omega_m+\omega_c -\omega_p-i\gamma_m/2}}
\eqname{EpsEIT}
\end{equation}
where $N(\x)$ is the atomic density and the parameter $f \approx
10^{-2}\gamma_e\lambda_{ge}^3$ is proportional to the oscillator
strength of the optical transition (of wavelength $\lambda_{ge}$) to which the weak probe beam of frequency $\omega_{p}$ couples. For $^{87}$Rb
atoms magnetically trapped into the sublevel $\ket{g}=\ket{F=2,M_F=2}$ of the $\textrm{S}_{1/2}$ ground state, the
probe couples $\ket{g}$ to an excited state $\ket{e}$ which can be a hyperfine component of the excited
$\textrm{P}_{1/2,3/2}$ states, while the coupling beam resonantly ($\omega_c=\omega_e-\omega_m$) couples the other
$F=1$ hyperfine component of the ground state, denoted by $\ket{m}$, to the same excited state $\ket{e}$. For this
choice of atomic levels, the $\ket{m}$ state dephasing $\gamma_m$ is orders of magnitude smaller than the excited
state decay $\gamma_e$ leading to a nearly complete suppression of probe absorption within a narrow linewidth
$4\Omega_c^2/\gamma_e$ around resonance $\omega_p=\omega_e-\omega_g$ and to a strong frequency
dispersion~\cite{EIT}. At resonance, the refractive index $\eta=\textrm{Re}\sqrt{\eps}=1$ and the
corresponding group velocity becomes,
\begin{equation}
v_{g}(\x)=\frac{c}{d [\omega_p \eta(\x,\omega_p)]/d\omega_p} \simeq
\frac{c\,\Omega_{c}^{2}}{2\pi f N(\x)\,
\omega_p}. \eqname{vg}
\end{equation}

\paragraph{\textbf{Transverse imaging}.}

\begin{figure}[htbp]
\begin{center}
\psfig{figure=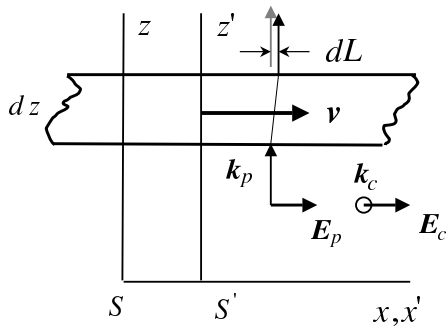,width=2.5in,angle=270}
\caption{Scheme for transverse imaging.} \label{fig:TFDscheme}
\end{center}
\end{figure}

A light ray beam propagating transversely across a slab of homogeneous
medium which slowly moves with constant speed $v$ along a direction parallel to its
boundaries (Fig.\ref{fig:TFDscheme} is seen to have a non--vanishing angle of incidence $v/ c$ in the medium rest
frame $S^{\p}$. The corresponding angle of refraction, obtained by using the Snell's law in $S^{\p}$ where the
medium is at rest, determines the common direction of phase and group velocity in $S^{\p}$. In the laboratory
frame $S$, however, phase and group velocities are no longer parallel: by using the fact that group velocities add
like particle velocities~\cite{Jackson} the group velocity in $S$ is found to be directed at an angle
\begin{equation}
\psi_{r}(\x) \simeq \frac{v}{v_g(\x)} - \frac{v/c}{\eta(\x,\omega_p)},
  \label{eq:psir}
\end{equation}
with respect to the $z$-direction. Appreciable bending of the ray beam may take place for group velocities not
much larger than $v$. This is seen to depend on the medium dispersion
through the light group velocity. Physically it arises from the motion
of the sample which acts 
as an effective transverse Fresnel drag onto
the beam of light that is then bent into the direction of motion~\cite{FresTr}.
Because the lateral shift subsequent to the deflection of the ray beam
inside the medium is directly proportional
to the velocity $v$, the drag effect can be reversibly exploited to image
the sample velocity.

The recent observation of ultraslow group velocities in cold~\cite{SlowVg} alkali atomic gases suggests that
substantial deflections may be observed in such media when undergoing collective motion at typical speeds of the
order of several recoil velocities~\cite{stringari}. Consider then a well focused probe beam propagating along $z$
through an atomic layer of thickness $dz$ transversally moving with velocity $v$ in the presence of a wide
coupling beam. The resonant coupling and probe beams are taken to be orthogonal to
the medium velocity so as to avoid any first order Doppler effect as shown in Fig.\ref{fig:TFDscheme}. For
simplicity we take the sample dynamics as consisting solely of a
uniform center of mass motion with a velocity
$v=v_{cm}$ along $x$~\cite{slosh-ketterle} and we assume that the current density $\textbf{J}(\x)=N(\x) \
\textbf{\textit{v}}$ remains essentially unchanged during the probe traversal time. The probe beam lateral
displacement, which is obtained from \eq{psir} with the help of \eq{vg} upon integrating over the whole sample
width in the $z$--direction,

\begin{equation}
L(x,y)\simeq
  \int\!\! \frac{v_{cm}}{v_{g}(\x)}\,dz
\simeq \frac{2\pi f \omega_p}{c\,\Omega_c^2}\,\int\!dz\,N(\x)\,v_{cm},
\label{eq:ld2}
\end{equation}
is just proportional to the column integral of the current density. By using slow--light parameters for which
$v_g$'s of the order of tens of m/sec are easily reached~\cite{SlowVg}, as well as realistic parameters for
Bose-Einstein condensates, where a $50\,\mu$m wide dense cloud of atoms moves at a speed of a few
cm/s~\cite{stringari}, we obtain a lateral shift \eq{ld2} of the order of a tenth of a $\mu$m. While the
measurement of such sub--micron or even smaller beam displacement is
well within the reach of current technology~\footnote{Standard
interferometric methods~\cite{jones} can be used as well as more
sophisticated two-quadrant split photodetectors~\cite{fabre}.}, the absence of absorption makes transverse
imaging a non-invasive scheme that is sound for {\em in-situ} observations of the atoms dynamics in a variety of
experimental situations involving collective modes of trapped atomic clouds such as, e.g., sloshing
oscillations~\cite{slosh-ketterle}. The nearly unit refractive index attained in the slow-light
regime also prevents significant image distortions due to lens--like refraction
effects~\cite{slosh-ketterle} at the condensate surfaces.

\paragraph{\textbf{Longitudinal imaging: vortices}}

A different scheme should be adopted when a spatially resolved image of the
velocity field within the cloud is
required. Consider a wide and resonant probe beam of wavevector $\kk_p$
incident on an atom cloud dressed by a
wide and resonant coupling beam of wavevector $\kk_c$~\cite{Ohberg}. The
propagation of the probe electric field
$\El_p(\x)$ can be described by using the usual wave equation for the
electromagnetic field~\cite{Jackson} with an
incident field in the form of a plane-wave as boundary condition. As source
term in the wave equation, the
dielectric polarization of the atoms has to be used. Because of Doppler
effect, the steep frequency dispersion of
the dielectric constant
  \eq{EpsEIT} implies a strong dependence of the dielectric
response of a moving EIT medium on its velocity. For a slow local velocity $\textit{\textbf{v}}(\x)$ and a
resonant probe, the medium is nearly transparent and its polarization
$\Pv(\x)$ is obtained by including in
\eq{EpsEIT} the Doppler shift of the coupling beam frequency as well as the spatial dependence of the probe field $\El_p(\x)$,\footnote{The usual
Doppler shift is recovered if the probe field $\El_p(\x)$ has a plane wave profile of wavevector $\kk_p$, in which
case the operator $\nabla_\x$ can be replaced by $i\kk_p$.}
\begin{equation}
\Pv(\x)=\frac{c}{2\pi\omega_{p}}\left[\frac{\textit{\textbf{v}}(\x)}{ v_{g}(\x)}
\cdot\left(\kk_c+i\nabla_\x\right)\right] \El_p(\x). \eqname{Polar}
\end{equation}
Since the dielectric polarization of the atomic sample is weak \footnote{For realistic values of the density and
of the flow velocity, the refractive index deviates from unity only by very small amounts of the order of
$10^{-3}$.}, we can approximate the atoms as responding to an unperturbed incident probe beam. Within this Born
approximation, the sample is well described by a refractive index proportional to the scalar product
$\J(\x)\cdot(\kk_c-\kk_p)$. Once the refractive properties  are known, the phase and the intensity
profiles of the transmitted probe electric field beam are completely determined by numerical integration of the
wave equation.


\begin{figure}[htbp]
\begin{center}
\psfig{figure=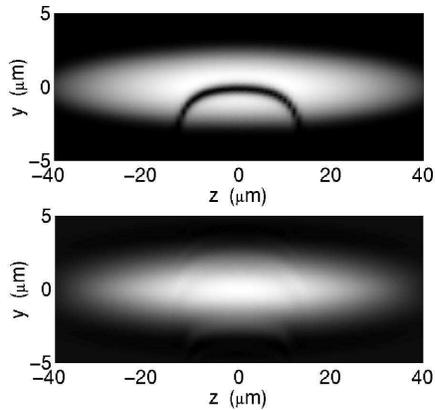,width=3.in}
\caption{Cross section  of the BEC density profile in the presence of a bent vortex (upper panel). Image of the
vortex taken by using conventional {\em in-situ} dispersive imaging
(lower panel): the dip in the density corresponding to the vortex core
can hardly be seen owing to diffraction effects. \label{fig:BentVortex}}
\end{center}
\end{figure}

Experimentally, the phase profile of the transmitted probe beam can be
determined by means of the same classical
phase reconstructing techniques~\cite{BornWolf} that have recently been
used to image the density profile of Bose
condensates. The {\em dark-ground} scheme~\cite{dark-ground-ketterle}
provides a picture in which the local
intensity is proportional to the square of the accumulated phase, while the
{\em phase-contrast}
scheme~\cite{disp-ketterle} gives a picture in which the intensity
variation is proportional to the accumulated
phase. Although the signal is stronger in the case of the phase contrast
picture, the dark--ground scheme has the
advantage of dealing with a zero measurement, in which no light is detected
for a vanishing velocity field.

\begin{figure}[htbp]
\begin{center}
\psfig{figure=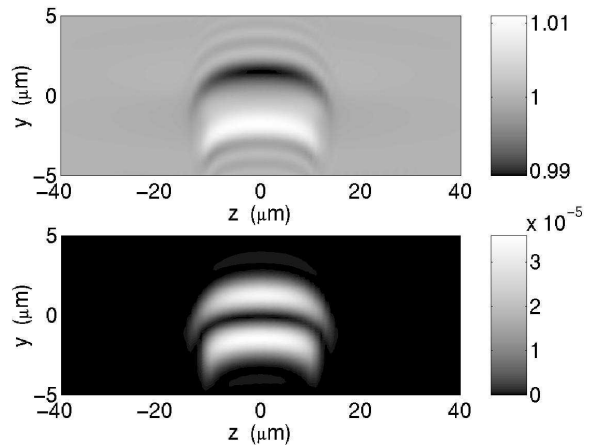,width=3.25in}
\caption{Phase-contrast (upper panel) and dark-ground (lower panel) images
of the bent vortex using slow light
imaging. The probe beam is taken along $x$ and the coupling along the
rotation axis $z$. The group velocity is
$v_g=1\,m/s$; the sample under consideration is the one considered in the
upper panel of Fig.2 of~\cite{BentVortex} }
\label{fig:BentVortex2}
\end{center}
\end{figure}
\begin{figure}[htbp]
\begin{center}
\psfig{figure=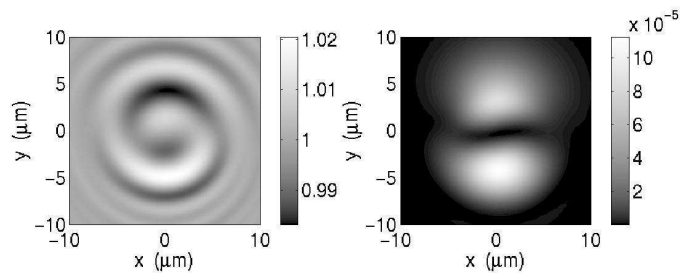,width=3.5in}
\caption{ Phase-contrast (left panel) and dark-ground (right panel) images of the bent vortex using slow--light
imaging. Same system as in Fig.\ref{fig:BentVortex2} except that the probe beam is here parallel to the rotation axis $z$
and the coupling is along $x$. \label{fig:BentVortexZ}}
\end{center}
\end{figure}
\begin{figure}[htbp]
\begin{center}
\includegraphics[width=2.75in]{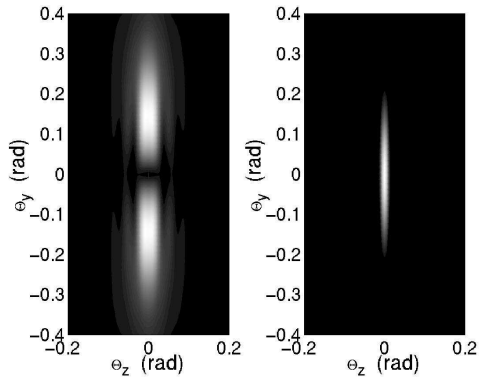}
\caption{Far-field diffraction pattern from a bent vortex using slow light imaging (left panel) and conventional
dispersive imaging (right panel) for the same situation as in Fig.\ref{fig:BentVortex2}. \label{fig:FarField}}
\end{center}
\end{figure}

We report in Figs.\ref{fig:BentVortex}-\ref{fig:FarField} results for the
specific case of a Bose-condensed cloud
containing a bent vortex~\cite{BentVortex,DalibBentVort} (upper panel of
Fig.\ref{fig:BentVortex}). Although our
numerical results fully take into account diffraction effects, a clear
interpretation of the resulting images can
be put forward by neglecting diffraction. Under this approximation, the
phase accumulated by the probe at a point
$(y,z)$ after crossing the atomic cloud is
\begin{equation}
\Delta \phi(y,z)=
(\kk_c-\kk_p)\cdot\int\!dx\,\frac{\vel(\x)}{v_g(\x)}
\eqname{dphi}
\end{equation}
For a coupling beam propagating along the rotation axis $z$ and a probe
propagating along the $x$ direction, the
accumulated phase has opposite signs on the two sides of the vortex core
from which stems the two-lobe structure
of the images in Fig.\ref{fig:BentVortex2}. For a condensate size much
larger than the core radius, the velocity
field around the vortex line has the typical $v=\frac{\hbar}{mr}$
behaviour.
For such a velocity field, the phase shift \eq{dphi} has a step-like
shape with constant and opposite values:
\begin{equation}
  \Delta \phi \approx \pm \pi \frac{\hbar \ | \kk_p |}{ m \ v_g}
\end{equation}
on either side of the vortex line
\footnote{For the parameters of Fig.\ref{fig:BentVortex2}, $\Delta \phi \approx 10^{-2}$ in good
agreement with the numerical calculation.}.
Contrary to the case of conventional dispersive imaging in which the
phase shift profile given by the decreased density at the
vortex core (whose diameter is
generally of the order of a fraction of $\mu
\textrm{m}$~\cite{slicing,DalibBentVort,CornellVort}) is strongly
affected by diffraction effects 
(cf. lower panel of Fig.\ref{fig:BentVortex}),
the phase shift in the slow light imaging case extends over the whole
condensate and therefore is much more robust against diffraction.
For this reason, a slow-light image of the vortex can be taken {\em in situ} without any
preliminary ballistic expansion stage.

As the local polarization is proportional to $\kk_p-\kk_c$, its magnitude
remains unchanged if the directions of
the coupling and probe beams are exchanged. Owing to the longer line of
sight for a probe propagating along the
rotation axis, the magnitude of the phase shift is expected to be somewhat
larger in this geometry. This is shown
in Fig.\ref{fig:BentVortexZ}, where  the coupling is taken along the $x$
axis and the probe is along the rotation
axis $z$: the $x$ component of the circulating current still gives the main
contribution to the phase--shift which
has opposite signs respectively above or below the $x$ axis. Unfortunately,
diffraction effects due to the rapid
transverse density variations increase the size of the spots and give rise
to fringes.

Finally, we would like to point out that in the case of slow--light
imaging, the presence of a vortex can also be
inferred from the far-field diffraction pattern~\cite{dark-ground-ketterle}
of the probe after scattering. As in
the Aharonov-Bohm-like picture, the presence of a circulation leads to the
appearance of fringes in the
diffraction pattern. We examine such a pattern in Fig.\ref{fig:FarField}.
As expected, the presence of the vortex
gives rise to a pair of symmetric side spots due to the interference
between the light passing from either side of
the vortex line and experiencing opposite phase-shifts. If a conventional
imaging technique sensitive to the
density were to be used instead, a single
spot without structure would be obtained and the fast density modulation
at the vortex core
would only result into an increased spot size.

The present results can clearly be extended to configurations with more than one vortex~\cite{slicing} or even to
multicomponent Bose-Einstein condensates exhibiting complex topological excitations~\cite{Monopoles}. The nearly
complete absence of absorption and the possibility of attaining images that are quite robust to diffraction make
slow--light imaging generally sound for the non-invasive observation of spatially small structures in ultracold
clouds of trapped atoms while their real-time dynamics can directly be followed {\em in situ}.

We are indebted to G. C. La Rocca and F. Bassani for fruitful suggestions
at various stages of the work and to M.
Modugno for providing us with the numerical results for the density and
current density profile of a bent vortex.
We also acknowledge stimulating discussions with M. Inguscio, S. Rolston,
Y. Castin, J.Dalibard, and S. Harris.
I.C. and M.A. acknowledge financial support from the EU (Contracts
HPMF-CT-2000-00901 and HPRICT1999-00111) as
well as from the INFM. Laboratoire Kastler Brossel is a unit\'e de Recherche de l'Ecole normale sup\'erieure et de l'Universit\'e
Pierre et Marie Curie, associ\'ee au
CNRS.

\end{document}